# Dome Craters on Ganymede and Callisto May Form by Topographic Relaxation of Pit Craters Aided by Remnant Impact Heat

M. L. Caussi[1] , A. J. Dombard[1] , D. G. Korycansky[2], O. L. White[3,4] , J. M. Moore[4] , and P. M. Schenk[5]

[1]Department of Earth & Environmental Sciences, University of Illinois Chicago, Chicago, IL, USA, [2]Department of Earth & Planetary Sciences, University of California, Santa Cruz, CA, USA, [3]SETI Institute, Mountain View, CA, USA, [4]NASA Ames Research Center, Mountain View, CA, USA, [5]Lunar and Planetary Institute (USRA), Houston, TX, USA

**Abstract** The icy Galilean satellites display impact crater morphologies that are rare in the Solar System. They deviate from the archetypal sequence of crater morphologies as a function of size found on rocky bodies and other icy satellites: they exhibit central pits in place of peaks, followed by central dome craters, anomalous dome craters, penepalimpsests, palimpsests, and multi-ring structures. Understanding the origin of these features will provide insight into the geophysical factors that operate within the icy Galilean satellites. Pit craters above a size threshold feature domes. This trend, and the similarity in morphology between the two classes, suggest a genetic link between pit and dome craters. We propose that dome craters evolve from pit craters through topographic relaxation, facilitated by remnant heat from the impact. Our finite element simulations show that, for the specific crater sizes where we see domes on Ganymede and Callisto, domes form from pit craters within 10 Myr. Topographic relaxation eliminates the stresses induced by crater topography and restores a flat surface: ice flows downwards from the rim and upwards from the crater depression driven by gravity. When the starting topography is a pit crater, the heat left over from the impact is concentrated below the pit. Since warm ice flows more rapidly, the upward flow is enhanced beneath the pit, leading to the emergence of a dome. Given the timescales and the dependence on heat flux, this model could be used to constrain the thermal history and evolution of these moons.

**Plain Language Summary** Ganymede and Callisto are large icy moons orbiting Jupiter, believed to be ocean worlds. These moons' surfaces display impact craters with shapes not seen elsewhere in the Solar System. This may be due to factors such as their high gravity, temperature, presence of subsurface oceans, and/or impactor characteristics. By exploring these atypical craters, we aim to shed light on the inner workings and evolution of these moons and, by extension, contribute to a clearer picture of the evolution of the outer Solar System. This study focuses on unveiling the origins of dome craters on Ganymede and Callisto, which have rounded bright domes in the center and are virtually unique to these moons. We believe these domes are connected to a relatively more common crater, the pit crater. Our simulations show that a pit crater can evolve into a dome crater within 10 million years as the ice flows slowly under its own weight. This flow is channeled back into the center creating a dome, aided by warm temperatures left over from the impact that soften the ice below the pit. This occurs only for the specific sizes where we see dome craters on these moons.

## 1. Introduction

The icy Galilean satellites contain an unparalleled variety of impact crater morphologies. Among these impact morphologies are pit and dome craters, which are rare elsewhere in the Solar System. Understanding the formation of these intriguing impact features can give us insight into the unique combination of geophysical factors that operate within the icy Galilean satellites. Intrinsic factors such as high surface gravity, rheology, thermal history and evolution as well as extrinsic factors such as impactor characteristics could play a role in the distinctive diversity of impact crater morphologies. In this paper, we focus on the origin of dome craters.

The icy Galilean satellites see a deviation from the archetypal sequence of crater morphologies as a function of size found on rocky bodies (like the Moon, Mars, and Mercury) and other icy outer Solar System satellites. As we examine larger craters, we typically see a transition from simple bowl shapes, to complex craters that exhibit central peaks, to peak-ring craters, and finally to multi-ring basins. In contrast, on Ganymede and Callisto, we see a transition to central pits in place of peaks, and the sequence continues with central dome craters, anomalous dome craters, penepalimpsests, palimpsests, and multi-ring structures (Schenk et al., 2004).

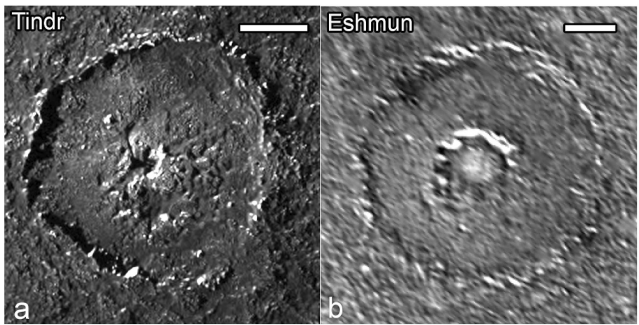

**Figure 1.** (a) Central pit crater Tindr on Callisto (2.3°S, 4.5°E) (b) Central dome crater Eshmun on Ganymede (17.5°S, 167.9°E). North is toward the top of the image; scale bars measure 20 km (White et al., 2022).

Central pit craters are predominant in the size range of ~35–~60 km in diameter (Schenk et al., 2004) and consist of craters with pits in their center (Figure 1a). Starting at ~60 km (Schenk et al., 2004), most of these central pits host a dome (Figure 1b). The emergence of dome craters at sizes where pit craters start to be less common and their similarity in morphology suggest that there may be a genetic link between the two (Bray et al., 2012; Schenk, 1993). We propose here that pit craters evolve into dome craters via topographic relaxation.

### 1.1. Pit and Dome Crater Morphology

Central pit craters are the predominant class for craters between ~35 and ~60 km in size on Ganymede and Callisto, and are characterized by a small rimmed pit in the center. Pit-to-crater diameter ratios increase with crater size, and crater depths remain constant or even decline with increasing crater size; a profile of pit crater Tindr on Callisto is shown in Figure 2a. Pit craters have been observed on Ceres and to some extent on Mars, but are rare or absent in the rest of the Solar System (Schenk et al., 2004).

Dome craters are predominant at diameters between ~60 and ~175 km and are morphologically similar to pit craters, but they exhibit a dome emerging from the floor of the rimmed central pit. Both pit and dome craters are shallower than would be expected for complex craters on Ganymede and Callisto (Schenk, 1993). Domes are large, rounded structures that can be up to 1.5 km high (Schenk et al., 2004), and can present a network of narrow fractures (Figure 3). Dome craters can be divided into two classes: central dome craters and anomalous dome craters.

Compared to anomalous dome craters, central dome craters tend to be smaller in size and younger with quite narrow sharp rims. The dome-to-diameter ratio increases with crater size, and rim-to-floor depths typically range from 0.8 to 1.6 km (Schenk et al., 2004). A profile of the dome crater Osiris on Ganymede is shown in Figure 2b.

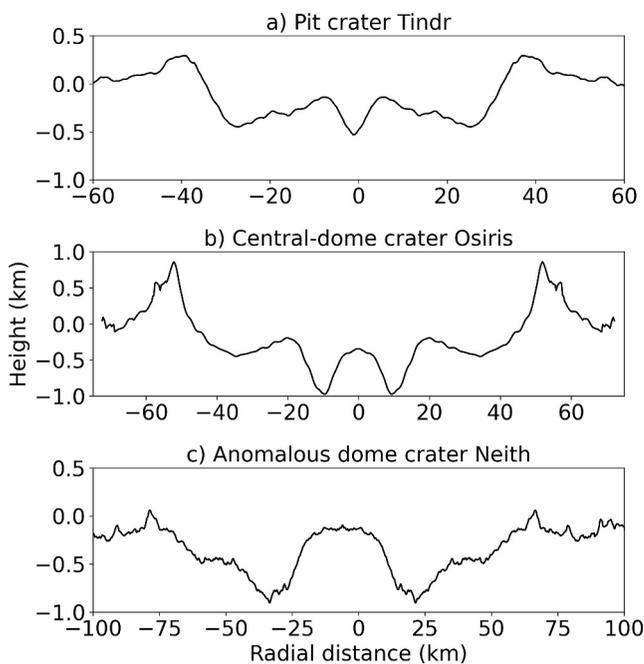

**Figure 2.** (a) Topographic profiles of Tindr pit crater, (b) Osiris central dome crater, and (c) anomalous dome crater Neith. Profiles were extracted using a photoclinometric digital elevation model (White et al., 2022).

Some very young dome craters such as Osiris are covered by a uniform bright frost deposit (Figure 3a). In older dome craters, this deposit has been erased, leaving a ~10% albedo contrast between the dome and the surrounding crater floor (Schenk, 1993).

Anomalous dome craters (Figure 3c) are similar in morphology to central dome craters, but they differ in some aspects: their rims do not form sharp ridges but rather tend to be broad, low and subdued rises -or, in some cases, they may be non-existent. The dome-to-diameter ratio is constant for all sizes, with wider domes compared to central dome craters. These features have also been called "large dome craters" by Moore and Malin (1988). Rim-to-floor depth is negligible or very shallow (Schenk et al., 2004), as shown in the topographic profile in Figure 2c. This crater class seems to be older due to the amount of superimposed craters and the lack of bright floor deposits (Schenk et al., 2004; White et al., 2022). Anomalous dome craters have been identified on Ganymede spanning a size range of 50–175 km in diameter, and of 50–250 km in diameter on Callisto (Schenk et al., 2004). The transition to anomalous dome morphology seems to have shifted over time from ~60 km, to ~150 km at present (Schenk, 1993).

### 1.2. Previous Hypotheses for Pit and Dome Origin

With increasing crater size, we observe a morphological transition from central peaks to central pits, and at larger dimensions, domes appear within pits. Therefore, the origin of dome craters may be tied to the origin of the pits themselves. Here, we review the past literature on the origin of both.

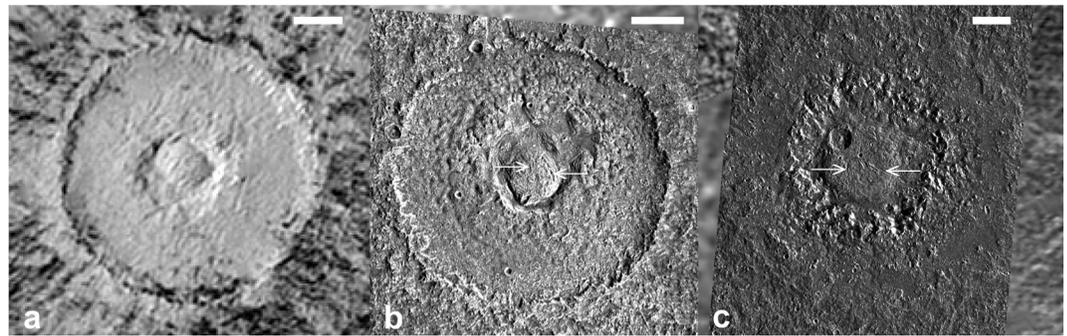

**Figure 3.** Dome craters on Ganymede (White et al., 2022); fracture networks can be seen at the surfaces of the domes (indicated with white arrows). (a) Central dome crater Osiris (38.0°S, 166.3°W): a uniform bright deposit covers the crater up to the rim. (b) Dome crater Melkart (9.9°S, 173.9°E) (c) Anomalous dome crater Neith (29.5°N, 7.0°W). North is toward the top of the images; scale bars measure 20 km.

### 1.2.1. Pit Origin

Pit craters have been hypothesized to originate via vapor explosions (e.g., Wood et al., 1978), peak collapse (e.g., Croft, 1983; Melosh, 1982), and melt drainage (e.g., Elder et al., 2012).

The vapor explosion hypothesis (e.g., Wood et al., 1978) was proposed for Martian pit craters, and states that upon impact, an explosive release of volatiles occurs during rebound, removing the core of the central peak and leaving a central pit. However, on Ganymede and Callisto, there is no evidence of the expected debris around the pit to support this hypothesis (Schenk et al., 2019).

The hypothesis of peak collapse (e.g., Croft, 1983; Melosh, 1982) posits that pits form during the impact modification stage when the central peak collapses due to the material rheology. However, it is unclear how a collapse would lead to a depression and not simply a disaggregated peak sitting in the center.

The hypothesis of melt drainage (e.g., Elder et al., 2012) proposes that the melt originating during impact can drain into the subsurface and leave a depression. This hypothesis is supported by the fact that pits occur in icy targets or those with some ice or volatiles, and not in purely rocky targets. In addition, pit crater formation seems to be favored in warmer temperatures (Schenk et al., 2019) where more melt can form. Moreover, melt volumes obtained from impact simulations (Elder et al., 2012; Korycansky, White, et al., 2022) are sufficient to create pits of the sizes found on Ganymede and Callisto if drained. We further build on this hypothesis in the discussion (Section 4.2).

### 1.2.2. Dome Origin

Diverse models for dome formation were proposed based on Voyager data. These included ice volcanism (Squyres, 1980), freezing and expansion of an impact-melt lake during crater formation (Croft, 1983), diapirism of low-density material mobilized post impact (Moore & Malin, 1988), and the central uplift of a layer of rheologically weak ice during impact (Schenk, 1993).

The icy volcanism hypothesis proposed in Squyres (1980) posits that two large domes located on grooved terrain originated when water from a shallow liquid mantle erupted to the surface through an impact fracture network. Current understanding of the thermal structure of Ganymede (e.g., Schubert et al., 1996) does not support the existence of such a shallow liquid mantle, as depths to the ocean are estimated to be of the order of 100 km. Additionally, it is unclear how a low-viscosity melt could construct a pronounced topographic dome.

The hypothesis of an impact-melt lake origin (Croft, 1983) has dome craters forming when a higher-than-average velocity impact creates a large melt volume. In these conditions, a melt pool in the center refreezes and expands to form a dome. Under this hypothesis, the dome crater is a direct product of the impact, with no role for relaxation or other post-impact processes. Following the acquisition of Galileo's higher-resolution observations, the melt-lake hypothesis began to be disfavored as no indications of flooding were found (Schenk et al., 2004).

The diapir hypothesis (Moore & Malin, 1988) proposes that a diapir rising in the center of the crater forms a dome. The buoyancy-driven upwelling is caused by a low-density layer at depth that is disturbed by the impact. After the impact, the upwarping of the layers below the central region of the crater generates a density instability, causing a diapir to rise and form the dome. Upwarping below the crater center is caused either by rebound upon impact or by post-impact relaxation flow. Once it reaches the surface, the diapir cools down and becomes more rigid, retaining its topography.

This hypothesis is consistent with the high albedo of the domes because the diapir would be compositionally distinct and cleaner than the surrounding ice. It is also consistent with the increase in dome diameter as craters get larger because the impact would be sampling from a low-density layer at constant depth (larger craters would excavate a larger portion of the layer). In contrast, post-impact diapirism is not consistent with undisturbed bright deposits observed in young dome craters: a diapir rising after the deposit formed could disturb it. This factor led the hypothesis to be disfavored (Schenk et al., 2004); however, whether the bright deposit covers the dome will be better resolved upon a return to Jupiter (Schenk et al., 2019). Moreover, because domes are found everywhere across Ganymede and Callisto's surfaces, a global layer that is gravitationally unstable (i.e., less dense than layers above it) would be required in both planetary bodies for this model to work, which is unlikely given these two moons' disparate geologic histories.

The hypothesis of central uplift upon impact was proposed by Schenk (1993) and consists of dome formation during the modification stage of the impact process. The upward flow of material into the center of the crater driven by collapse causes a dome to emerge in a manner similar to peak emergence for terrestrial and lunar complex craters. The pit walls would be the scarps from the broken upper layer that is mechanically stiff. This hypothesis requires that the uplifted material is rheologically weak to explain the smooth and rounded appearance of the dome, which contrasts with that of a peak.

In the same manner as the previous hypothesis, the increase in pit diameter with increasing crater diameter is explained by the sampling of a layer at constant depth. In addition, the albedo differences observed between domes and adjacent terrain are explained by the compositionally distinct dome-forming layer. This hypothesis also addresses the undisturbed bright frost deposits: because of the short timescales involved, the dome would form before the emplacement of the bright deposits, leaving these undisturbed. However, this hypothesis demands a rheologically weak global layer in both planetary bodies; in addition, it is still unclear if the impact mechanics leading up to the dome would work in this way. Numerical modeling would be required in order to learn more about impact mechanics in these conditions. Recent hydrocode modeling of impacts into icy targets like Ganymede and Callisto (e.g., Korycansky, Caussi, et al., 2022) has rendered shallow craters with large melt pools but no central features (see below).

More recently, the similarity in morphology of Ceres' Occator crater to dome craters on Ganymede and Callisto prompted a comparison (Schenk et al., 2019). This crater, unique to Ceres, presents a dome with similarly disposed fractures situated within a pit. The formation sequence proposed in Schenk et al. (2019) based on stratigraphic relationships is that the pit originates first, and the dome forms afterward via a process that inflates the surface from below and causes the crest of the dome to fracture. Quick et al. (2019) considered cryovolcanism as a potential mechanism driving this process, hypothesizing a preexisting cryomagma chamber beneath the Occator crater. However, certain discrepancies between the dome crater on Ceres and those on Ganymede and Callisto challenge the notion of a shared formation mechanism. For instance, the diameter of the Occator dome is smaller than that of the pit floor, deviating from the patterns observed on Ganymede and Callisto, where the dome fully, or mostly, encompasses the pit. Furthermore, Ceres is a rocky world with an ice component, which could lead to a distinctive dome formation process compared to that on bodies with surfaces that are mainly icy. Finally, the dome at Occator is the only one of its kind on Ceres, suggesting a region-specific variance that enabled the formation of the dome.

### 1.3. Topographic Relaxation Model

Here, we explore the idea that domes may originate by topographic relaxation of an initial central depression in a pit crater, evolving it to a central dome crater over longer, post-impact time scales. Topographic relaxation occurs more rapidly on icy terrain relative to rock due to its generally lower viscosity. It consists of the slow flow (creep) of ice in response to stresses induced by topography. To relax the induced stresses and restore a flat surface, ice flows downwards from the rim and upwards from the crater depression and the pit over long timescales.

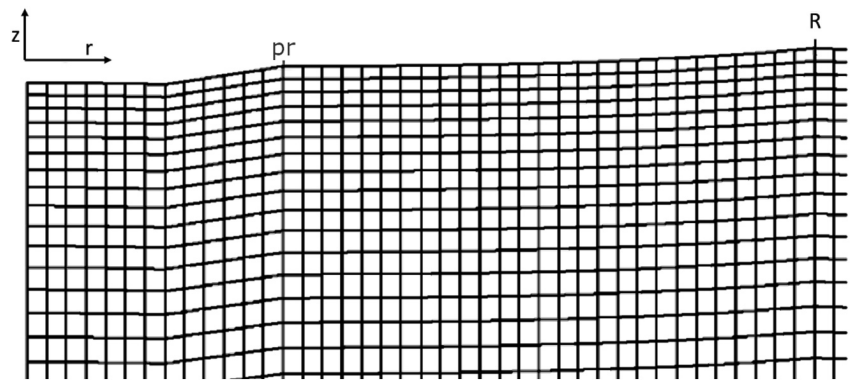

**Figure 4.** Example mesh used for Osiris-sized crater simulations ($R = 54$ km). The mesh depicted here is truncated for visualization purposes, extending to a distance $R$ from the center of the crater; the pit radius ($pr$) is shown for reference. The actual domain of the mesh extends up to 3 $R$ in both the horizontal dimension ($r$) and the vertical dimension ($z$).

Viscous forces resist creep flow; therefore, the lower the viscosity of the material, the more rapid the creep. Because viscosity decreases with increasing temperature, relaxation is enhanced if the ice is warm (e.g., Dombard & McKinnon, 2006a). We consider two sources of heat affecting relaxation: the background heat flux from the satellites and the heat left over from the impact itself. The mechanism of topographic relaxation with the addition of the remnant impact heat has been used to explain the extreme elevation of some central peaks above the background terrain on Saturn's moon Dione (Dombard et al., 2007): a thermal anomaly in the center of the crater, left over from the impact, can enhance the creeping flow of ice, enhancing the uplift of the center of the crater on which the central peak lies. A similar mechanism where the central pit is uplifted into a dome might explain central dome craters.

## 2. Methods

The viscoelastic relaxation of dome craters is investigated using finite element modeling. This method consists of subdividing the domain into discretely sized subregions (finite elements) and numerically solving the governing differential equations of heat transfer and displacement for each element. These simulations are conducted using the commercially available Hexagon Marc package, which has been used previously to explore relaxation on these moons (e.g., Dombard & McKinnon, 2006a). We simulate the evolution of pit craters over 1 Gyr using the same basic method benchmarked by Dombard and McKinnon (2006a). First, a thermal simulation that tracks the diffusion of the impact heat is performed, the results of which are then mapped into a mechanical simulation that solves the long-term evolution of the topography.

### 2.1. Mesh

The simulation space is one radial plane with a size of three times the radius ($R$) of the crater in both the horizontal and vertical directions. The mesh subdivides the domain into 7,200 quadrilateral elements with 120 divisions along the $z$ axis and 60 divisions along the $r$ axis. Each element has four nodes on the vertices and is solved with bilinear shape functions. The center of the crater is at $r = 0$, and the mesh is axisymmetric around the $z$ axis (see Figure 4). A bias factor -defined as a parameter to skew the density of elements toward areas of higher interest- is set to 0.35 to concentrate the elements closer to the top of the mesh where most of the deformation is happening. A length-to-width aspect ratio is kept under 4:1 for finite element stability (Akin, 1994).

### 2.2. Initial Crater Shape

The initial crater shape is that of an unrelaxed pit crater. The shape of the main crater depression (bowl) is a fourth order polynomial, while the slope of the ejecta blanket that extends beyond the rim is modeled following an inverse third power law (Dombard & McKinnon, 2006a), with its thickness fading off completely at two crater radii. The pit is modeled with straight walls and a small dome on the pit bottom shaped as a second order polynomial; in some cases, a flat bottom is used (see Table 1). We will argue in the discussion section why we include a small dome at the bottom of the pit at the start of our simulations. Vertical distances (see Figure 5) such as the rim height (*rh*) and crater depth (*d*) are estimated based on trends for fresh pit craters on these worlds (Bray et al., 2012). Horizontal dimensions such as the radius of the central feature (*cfr*), the pit radius (*pr*), and the crater radius (*R*) shown in Figure 5 are taken from Schenk et al. (2004) and White et al. (2022). Table 1 shows the specific horizontal and vertical distances used for each of the simulations.

**Table 1**
*Simulation Parameters and Summary*

| Diameter (D) | Original shape (description) | Vertical dimensions (km) | | | | Horizontal dimensions (km) | | | Thermal state | Dome formation (Yes/No) |
|---|---|---|---|---|---|---|---|---|---|---|
| | | Initial dome height (*cfh*) | Pit depth (*pd*) | Crater depth (*d*) | Rim height (*rh*) | Central feature radius (*cfr*) | Pit radius (*pr*) | Crater radius (*R*) | | |
| 145 km (Neith-size) | Neith-sized pit crater with small "seed dome" | 0.15 | 1.5 | 1.4 | 0.67 | 27 | 41 | 72.5 | $q = 3$ mW m$^{-2}$, with thermal anomaly | Yes |
| | | | | | | | | | $q = 10$ mW m$^{-2}$, with thermal anomaly | No |
| 108 km (Osiris-size) | Osiris-sized pit crater with small "seed dome" | 0.060 | 1.2 | 1.2 | 0.67 | 9.0 | 17 | 54 | $q = 3.5$ mW m$^{-2}$, with thermal anomaly | Yes |
| | | | | | | | | | $q = 10$ mW m$^{-2}$, with thermal anomaly | Yes |
| | | | | | | | | | $q = 3.5$ mW m$^{-2}$, no thermal anomaly | No |
| | Dome crater (present-day Osiris topography) | 0.65 | 0.55 | 1.3 | 0.8 | 9.0 | 17 | 54 | $q = 3$ mW m$^{-2}$, with thermal anomaly | Yes |
| 58 km (overlapping pit/dome crater size range) | Pit crater with narrow central pit | 0 | 0.50 | 1.0 | 0.60 | 2.5 | 9.0 | 29 | $q = 3$ mW m$^{-2}$, with thermal anomaly | No |
| | | | | | | | | | $q = 10$ mW m$^{-2}$, with thermal anomaly | No |
| | Pit crater with wide central pit | 0 | 0.50 | 1.0 | 0.60 | 6.5 | 10 | 29 | $q = 3$ mW m$^{-2}$, with thermal anomaly | No |
| | | | | | | | | | $q = 10$ mW m$^{-2}$, with thermal anomaly | Yes |
| 40 km (pit-crater size range) | Pit crater | 0 | 0.30 | 0.87 | 0.46 | 2.0 | 3.5 | 20 | $q = 3$ mW m$^{-2}$, with thermal anomaly | No |
| | | | | | | | | | $q = 10$ mW m$^{-2}$, with thermal anomaly | No |

### 2.3. Thermal Simulations

Because the instantaneous temperature field strongly influences the viscosity and, hence, the rate of relaxation, a thermal simulation is performed first to determine the subsurface temperature evolution as the remnant heat from the impact diffuses.

A conservative estimate of the background heat flux is applied to the elements at the base of the mesh, with a value of 3 mW m$^{-2}$ selected for the heat flow (Bland et al., 2009; Mueller & McKinnon, 1988) expected in the present epoch, or 10 mW m$^{-2}$, representing earlier conditions. These values of heat flux are consistent with the relaxation results of Dombard and McKinnon (2006a), and the more recent ones of Bland and Bray (2024). The sides of the mesh are set to zero heat flux, and the surface temperatures are fixed at 120 K. We apply a thermal conductivity of ice that follows an inversely proportional relationship with temperature and equals 651/T (Petrenko & Whitworth, 1999). The thermal solution requires diffusivity, which is a combination of conductivity, mass density, and specific heat. We assume that variability in diffusivity arises from the conductivity, and we use a value of specific heat for ice of 894 J/KgK, corresponding to a surface temperature of 120 K (Melinder, 2010). The thermal simulations output the temperature structure throughout a period of a billion years. These outputs are input into the mechanical simulations in order to solve the evolution of the topography throughout the same time period. A coupled thermal-mechanical analysis is computationally expensive and not necessary for this case where the deformation does not alter the thermal structure significantly (see Dombard & McKinnon, 2006a).

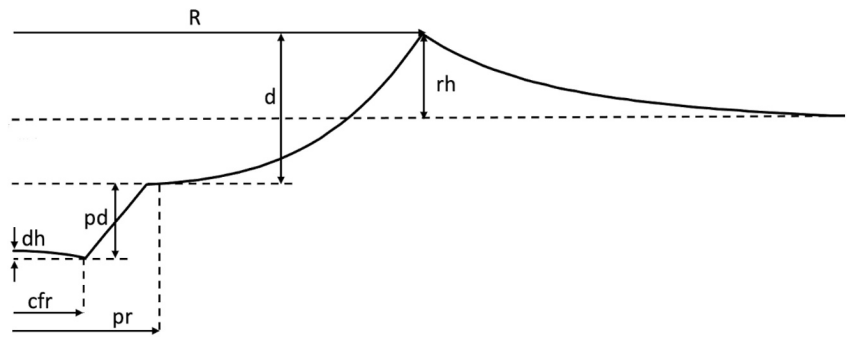

**Figure 5.** Dimensions used for modeling the initial geometry.

### 2.3.1. Thermal Anomaly Setup

The dimensions and magnitudes of the impact-induced thermal anomaly are constrained from impact simulations (Korycansky, Caussi, et al., 2022). These are iSALE hydrocode simulations that track the pressures reached in the impacted ice in order to calculate the volume of melt that originates from the impact. The areas that were shocked to the relevant pressures will exhibit some degree of melting. The final results of the simulations showed a melted area that is generally an order of magnitude larger than the pit volumes inferred from surface topography, with most of the melt concentrated below the center of the crater (Figure 6).

Our simulations do not include phase changes, so we start them at the point where all the melt has solidified, setting the nodes to a temperature of 273 K in a spatial pattern guided by the hydrocode results.

In our simulations, we use a smaller melt area that corresponds to the radius of the pit. This approach aligns with our hypothesis for pit formation through melt drainage, which we will discuss in Section 4.2. Reducing the size of the thermal anomaly reduces the rate of relaxation at the center of the crater compared to using a more realistic approximation to the temperature profiles in Korycansky, Caussi, et al. (2022), where the melt pool is larger; however, the implementation in Marc is less time consuming and it does not fundamentally change the results. The thermal anomaly in our simulations is located immediately below the surface nodes, its lateral extent is that of the pit floor, and its vertical extent is roughly that of the pit floor radius.

### 2.3.2. Thermal Simulation Structure

The purpose of the thermal simulations is to track the conductive dissipation of the impact heat by generating the temperature profiles at each point in time by numerically solving the heat diffusion equation. To accomplish this task, the finite element package first solves for the steady-state temperature profile of a half space subjected to an edge heat flux, fixed surface temperature, and a fixed thermal anomaly. The equilibrium temperatures from the steady-state solution produce a thermal profile that is used as the starting point for solving the transient heat diffusion equation, where the anomalous high temperatures are no longer fixed and therefore dissipate over time (see Figure 7).

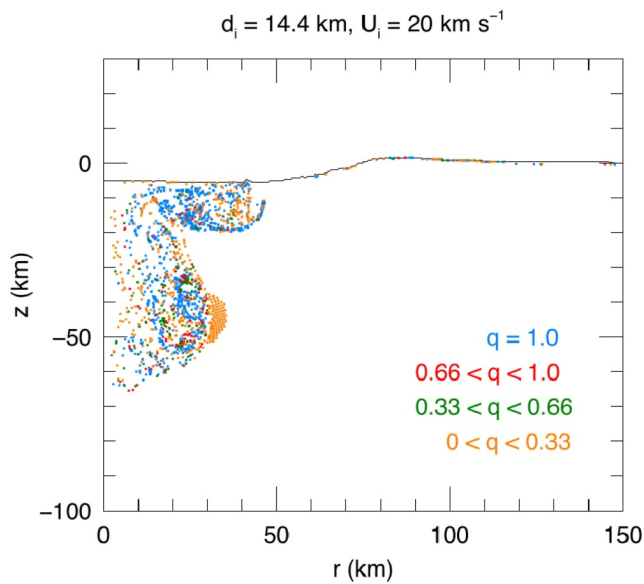

**Figure 6.** Melt originated from a vertical impact into an icy target at $T = 125$ K, with an impactor size of 14.4 km in diameter, and a velocity of 20 km s$^{-1}$ (Korycansky, Caussi, et al., 2022). The fraction of partial melting is denoted by $q$.

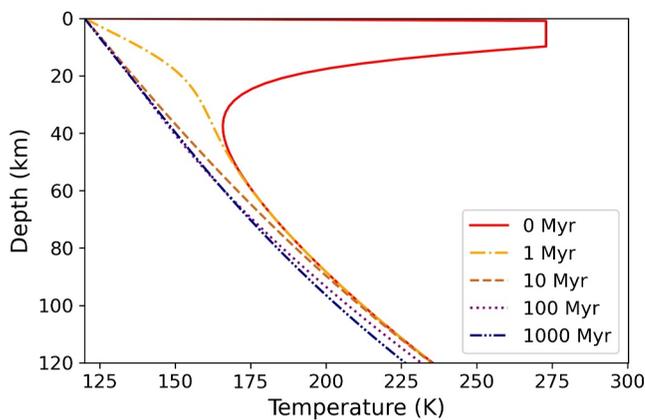

**Figure 7.** Temperature profile evolution over 1 Gyr for a 108-km pit crater under a heat flux of 3.5 mW m$^{-2}$, with the profile taken along the axis of the crater. The thermal anomaly left over from the impact at shallow depth dissipates roughly within 10 Myr.

### 2.4. Mechanical Simulations

Mechanical simulations were performed using the results from the thermal simulations. A uniform gravity load is applied to all elements in the mesh, with a gravitational acceleration of 1.44 m s$^{-2}$ corresponding to Ganymede. Gravity is what generates the overburden stresses that drive relaxation. Motions on the side boundaries are restricted to free slip (displacement can occur only along the direction of the boundary), and the bottom boundary is fixed so that displacement cannot occur in any direction. The density of the material is set to be 950 kg m$^{-3}$, corresponding to hexagonal water ice with zero porosity at a temperature of 120 K and a pressure of ∼110 MPa, which exists at a depth of ∼80 km on Ganymede, and of ∼90 km on Callisto (Dombard & McKinnon, 2006a). Large-strain formulation is used in the simulations, which includes the second-order terms in the strain-displacement relationships (see Ranalli, 1995, Section 2.8). The displacements are not large, but the large-strain formulation in Marc incorporates a geometric update, where the stresses are recalculated as displacement occurs. Because the driving stresses are a result of the topography, neglecting the geometric update means that the stresses would always be based on the initial topography, artificially enhancing the rate of relaxation. Deviatoric strain at each element is computed using four integration points and bilinear shape functions, while volumetric strain (dilatation) is computed using one-point integration with constant dilatation throughout the element, which helps prevent possible volumetric locking of the elements.

### 2.5. Rheology

The material behavior of ice is modeled as a Maxwell viscoelastic solid. These materials accommodate stresses with a combination of elastic and viscous strains, the total strain being the sum of the two:

$$\varepsilon_{ij}^{\text{total}} = \varepsilon_{ij}^{\text{elastic}} + \varepsilon_{ij}^{\text{viscous}} \tag{1}$$

Maxwell viscoelasticity is commonly represented by an elastic spring linked in series to a viscous dashpot, where the spring deforms instantaneously in response to a load and the dashpot deforms over time. The timescale over which viscous deformation in the dashpot dominates (i.e., Maxwell time) increases with viscosity. Thus, the dominant response to stress in colder and stiffer materials tends to remain elastic for a longer period of time than for materials with lower viscosity. Low-viscosity materials, where the dashpot is the weak link, exhibit predominantly viscous behavior sooner.

The elastic behavior dominates in the colder and stiffer ice near the surface, constituting the lithosphere. The lithosphere bends to restore a flat surface when loaded with crater topography, flexing upwards on the crater depression and downwards under the rim. The viscous behavior dominates in the warmer and softer ice at depth, which creeps downwards from the rim and upwards from the depression, allowing the lithosphere above to flex. Given enough time, the bottom of the lithosphere starts to creep, thinning the lithosphere and allowing it to bend even more (e.g., Damptz & Dombard, 2011). The bending stresses in the flexed layer can exceed those imposed initially by the topography, enhancing the rate of relaxation when compared to purely viscous materials (see Dombard & McKinnon, 2006a). The process continues until all stresses in the upper layer are relaxed. In contrast to a purely viscous substrate where there is no instantaneous deformation and the driving stresses can only decrease over time, a viscoelastic substrate relaxes faster both because there is instantaneous elastic deformation and because the driving stresses can increase over time. See Dombard and McKinnon (2006a) for a more extensive description of viscoelastic relaxation.

When the bending stresses exceed the yield strength of the ice, the lithosphere, which is the coldest and stiffest layer, behaves brittlely and exhibits fracture. However, plasticity is not incorporated in our model due to a lack of resolution in the dome where fractures would concentrate. Plasticity would only increase the rate of relaxation by undermining the strength of the flexural layer, although previous work (Dombard & McKinnon, 2006a) has shown that plasticity is only a minor contributor to relaxation.

### 2.5.1. Elastic Parameters

The elastic deformation is modeled using linear isotropic elasticity with a Young's modulus of 9.33 GPa and a Poisson's ratio of 0.33 (Gammon et al., 1983). However, if ice were modeled as compressible as determined by this Poisson's ratio value, it would be compacted by its own gravity in the vertical direction at the start of the simulations even in the absence of topography: the overburden stresses in the vertical direction would be larger than the stresses in the horizontal direction, and this imbalance would cause deviatoric stresses (see Turcotte & Schubert, 2014, Section 3.4, uniaxial strain). The state of stress in the subsurface of a planetary body with no topography, and in the absence of other forces, is assumed to be hydrostatic (Heim's rule) because deviatoric stresses would have been eliminated due to creep given sufficient time (Jaeger et al., 2007). Therefore, to have the state of stress in the subsurface be hydrostatic and perturbed by crater topography, the ice is set to be virtually incompressible (Poisson's ratio very near 0.5). In order to compensate for this adjustment, we maintain the original flexural rigidity (see Turcotte & Schubert, 2014, eq. 372). Because the lithosphere can be considered as an elastic layer bending in response to loading, the elastic portion of the relaxation is mainly controlled by its ability to flex (Dombard & McKinnon, 2006a). In order to maintain the flexural rigidity with this adjusted Poisson's ratio, Young's modulus is set to be 7.82 GPa. This change in elastic parameters does not affect relaxation significantly, as demonstrated by Dombard and McKinnon (2006b).

### 2.5.2. Viscous Parameters

The viscous deformation portion is modeled using the ice rheology described by Goldsby and Kohlstedt (2001). Because of the long timescales, viscous flow is considered solely due to steady-state creep, assuming a grain size of 1 mm. This rheology entails a combination of creep mechanisms acting together, each one dominating over a particular temperature and stress range. The constitutive equation determined experimentally relates the creep mechanisms of grain boundary sliding (GBS), basal slip (ES), dislocation creep (dis), and diffusion creep (diff) to the total viscous strain rate ($\dot{\varepsilon}^{\text{viscous}}$):

$$\dot{\varepsilon}^{\text{viscous}} = \left(\frac{1}{\dot{\varepsilon}_{\text{GBS}}} + \frac{1}{\dot{\varepsilon}_{\text{ES}}}\right)^{-1} + \dot{\varepsilon}_{\text{dis}} + \dot{\varepsilon}_{\text{diff}}. \tag{2}$$

Basal slip and GBS are mutually accommodating mechanisms, with the slower one limiting the other, and thus act in parallel. Dislocation creep and diffusion creep act in series, adding to the total deformation.

Each mechanism can be modeled by a power law of the form:

$$\dot{\varepsilon}_{\text{eff}}^{\text{viscous}} = A \left(\frac{1}{\delta}\right)^m \sigma_{\text{eff}}'^n e^{-\frac{Q+PV}{RT}}, \tag{3}$$

where $\dot{\varepsilon}_{\text{eff}}$ is the effective deviatoric strain rate, $\sigma_{\text{eff}}'$ is the effective deviatoric stress, $A$ is the pre-exponential constant normalized for uniaxial tension, $\delta$ is the grain size, $m$ is the grain size index, $n$ is the power law index, $Q$ is the activation energy, $T$ is temperature, $R$ is the gas constant, $P$ is the pressure, and $V$ is the activation volume. The contribution from the activation volume term is minor compared to the activation energy and falls within the bounds of experimental uncertainties (see Durham et al., 1997), so we omit it from our simulations; the rest of the creep parameters are taken from Goldsby and Kohlstedt (2001).

The finite element package Marc is internally normalized for uniaxial creep, which allows the use of the pre-exponential constant $A$ directly from the literature on uniaxial creep experiments. In uniaxial creep experiments, a sample is deformed along one direction. This setup allows for the determination of a flow law, which establishes a relationship between stress and strain rate. The empirically determined pre-exponential constant in this flow law can be used directly in Marc without requiring any further adjustments. Other similar codes, such as Tekton, require the conversion of the pre-exponential constant to one corresponding to a triaxial state of stress (see Ranalli, 1995. p.75).

Diffusion creep, typically very slow and operative under very low stresses, was not successfully measured in the study by Goldsby and Kohlstedt (2001). Consequently, an empirically derived flow law could not be established. We incorporate diffusion creep by using the estimated diffusion creep flow rate in Goldsby and Kohlstedt (2001), which is obtained by constraining the parameters in the diffusion creep equation with experimental data.

### 2.5.3. Mechanical Simulation Structure

In the first instant (time zero) of the simulations, the gravity load must be balanced by the internal stresses in the material for the system to reach an initial mechanical equilibrium. This step involves a very large adjustment in the system in no time, and to avoid numerical errors, an initial elastic iteration at time zero is included. During this iteration, all strains are accommodated elastically because viscous flow requires time. After the time-zero elastic solution, the system has internal stresses that drive viscoelastic relaxation in subsequent iterations. The strain rates are determined by Equation 1 and then by Equation 2.

### 2.5.4. Time Stepping

To select the time increments in the mechanical simulations, Marc calculates the ratio of viscous strain to elastic strain at each step everywhere in the mesh, keeping the time increment small enough that the ratio does not exceed a tolerance based on a user-defined fraction of the Maxwell time. For our simulations, we use a factor of 0.5 so that the ratio of creep strain to elastic strain does not exceed half the minimum Maxwell time at any increment. Time steps smaller than half the Maxwell time are required for numerical stability in viscoelastic simulations and are required to resolve the ductile deformation.

The Maxwell time decreases when viscosity decreases. Because viscosity decreases with increasing temperature, points at the bottom of the mesh, where the warmest temperatures would be, have the shortest Maxwell times. If these Maxwell times get overly small, it could become too computationally expensive. To avoid this problem, we set a cutoff for viscosity, a minimum value that it cannot dip below. Physically, this cutoff could represent the melting point of ice or convection in the ice shell. This cutoff only affects the elements below those near the surface of the mesh (i.e., below the lithosphere), so it does not interfere with the near-surface flow field where most of the deformation is happening.

Different viscosity cutoff values were tested in Dombard and McKinnon (2006a), and they did not affect the outcomes significantly. What controls the amount of relaxation is the viscous deformation at the base of the lithosphere which allows for continued bending of the flexural layer. Initially, the main role of the bottom material in relaxation is to get out of the way to accommodate the near-surface flexure. After that, the rate at which creep at the base of the lithosphere occurs is what controls the rate of relaxation, because it is what allows further bending. After a given amount of time, solutions with different cutoff viscosities converge into a common amount of relaxation, indicating that the viscosity at the bottom of the mesh does not interfere with the final relaxation outcomes (see Figure 11, Dombard & McKinnon, 2006a). For our simulations, we used a viscosity cutoff of $1 \times 10^{18}$ Pa s.

## 3. Results

Our simulation results show that central pit craters can evolve into dome craters in ∼10–100 Myr via topographic relaxation aided by remnant heat from the impact. This phenomenon occurs for the size range that exhibits domes on Ganymede and Callisto. Crater sizes at the extremes of the dome-crater size range are also explored in order to reproduce the morphological transitions.

Our results are consistent with the morphological transitions, as we reproduce the pattern of pit craters occurring at smaller crater sizes (∼35–60 km) and dome craters occurring at larger sizes (∼60–175 km). Around the morphological transition between these classes (∼60 km), where both pit and dome craters overlap on Ganymede and Callisto, we find that domes emerge only if the central pit is wide, and the heat flux is high.

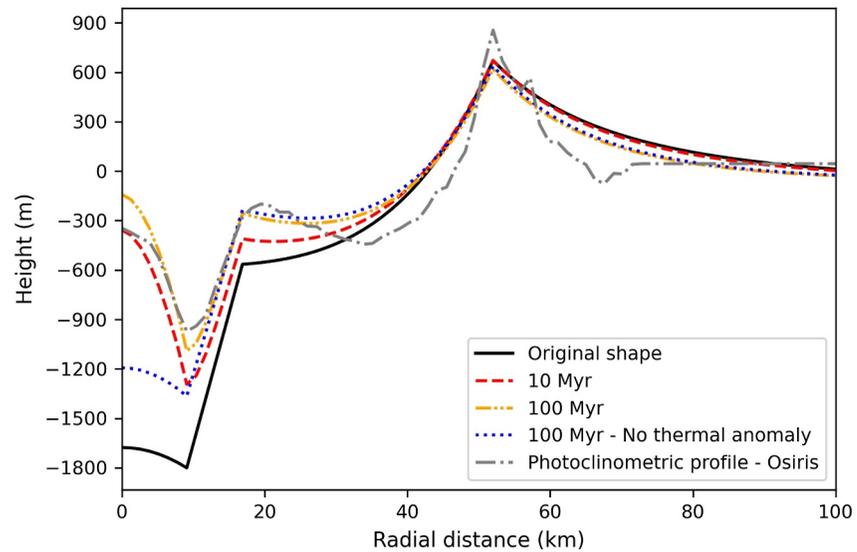

**Figure 8.** Topographic profiles of a relaxing 108-km pit crater under a heat flux of 3.5 mW m$^{-2}$. A dome is formed in 10 Myr when a thermal anomaly is included (red dashed line); further evolution to 100 Myr is shown in orange (double-dot dashed line). The observed topographic profile of Osiris crater as measured from a photoclinometric digital elevation model (White et al., 2022) is shown as a gray dot-dash line for comparison. A simulation using no remnant impact heat is shown in blue (dotted line). The assumed initial crater shape is shown in black (solid line).

### 3.1. Simulations Over the Dome-Crater Size Range (D ~60–175 km)

When a thermal anomaly left over from the impact is applied to pit craters in the dome size range, a dome is formed in ∼10 Myr. The thermal anomaly softens the ice below the pit, enhancing its upward flow as compared to the rest of the crater depression. In this size range, we performed simulations for craters with diameters of 108 and 145 km.

In the case of the 108-km-diameter crater, the resulting topographic profiles closely match that of Osiris (dashed gray line in Figure 8), a young dome crater on Ganymede. Notably, the dome forms while the thermal anomaly is largely extant. Figure 7 shows the temperature profiles along the central axis of the crater as the thermal anomaly dissipates over time. Comparing Figures 7 and 8, the thermal anomaly largely dissipates on time scales of order 1–10 Myr, enhancing the creeping flow of the ice under the pit when it is there: the dome profile obtained at 1 Myr is largely the same as that obtained at 10 Myr, in correspondence with the dissipation timescales of the thermal anomaly. After the anomaly dissipates, additional movement of the pit coincides with the basin as a whole.

Additionally, formation of a dome is always coincident with relaxation of the crater as a whole, because enhanced uplift in the central structure can only occur when the surrounding ice is soft enough to accommodate this extra flow. These conditions occur when the heat flow is high enough to allow relaxation of the crater as a whole (cf. Dombard et al., 2007). A low heat flux of 3.5 mW m$^{-2}$ yielded the best match with a profile of Osiris measured from a photoclinometric elevation model (White et al., 2022). Notably, the upbowing of the whole basin results in the creation of a subtly elevated pit rim, which is typically seen at dome craters.

When no thermal anomaly from the impact is applied (blue line in Figure 8), there is an overall uplift of the crater depression, but a dome does not form, indicating that there is no preferential uplift of the material below the pit. The profile of the relatively small-scale pit changes little and is essentially shifted upwards by the overall relaxation of the basin. When no thermal anomaly is applied but the heat flux is high (10 mW m$^{-2}$), the seed-dome height is enhanced and there is significantly more overall relaxation of the crater depression. Ultimately, it does not result in morphology similar to that of present-day Osiris, as the final dome height is very low relative to the pit depth.

Further evolution to 1 Gyr of the Osiris-sized crater is shown in Figure 9, both with a low heat flux of 3.5 mW m$^{-2}$ (Figure 9a), and a higher heat flux of 10 mW m$^{-2}$ (Figure 9b); the 10 and 100 Myr curves are the same as in Figure 8. Higher heat fluxes lead to taller domes because the whole subsurface is warmer and can accommodate the extra flow of the softer ice in the thermal-anomaly region more efficiently.

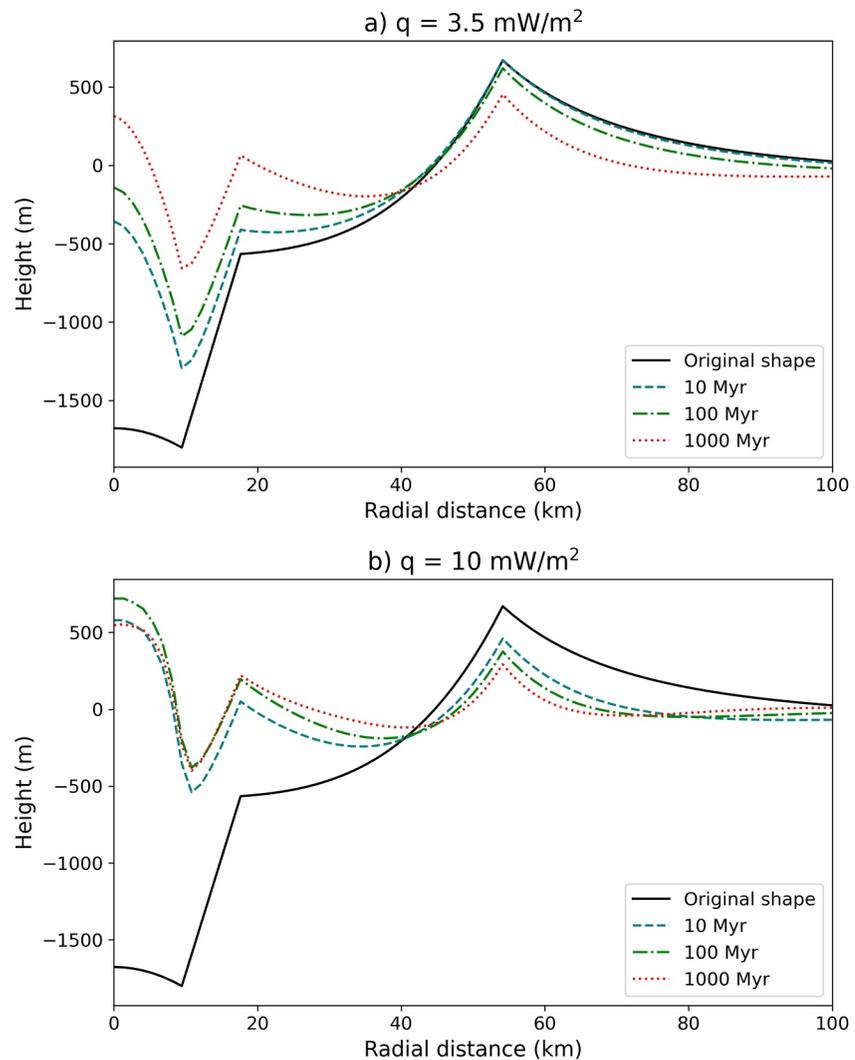

**Figure 9.** Topographic profiles of a relaxing 108-km pit crater under a heat flux of (a) 3.5 mW m$^{-2}$ and (b) 10 mW m$^{-2}$, with a thermal anomaly included for both cases. Higher heat flux leads to taller domes.

For the 145-km size crater, simulations under low heat flux (3 mW m$^{-2}$) and high heat flux (10 mW m$^{-2}$) are performed (Figure 10), where a thermal anomaly is included. For low heat fluxes of 3 mW m$^{-2}$, a dome is formed within 10 Myr. The photoclinometric profile of Neith, an anomalous dome crater on Ganymede, is shown for comparison (dashed gray line in Figure 10a). The outer bound of Neith's outer rim has a diameter of 170 km, but the rim is broad and its height tends to peak at ∼145 km; for the purpose of these simulations, we use 145 km as Neith's diameter. The simulation results show a sharp corner at the dome flanks that is likely an artifact of the short-wavelength sharp corner in the original shape; it does not match the rounded appearance of the dome in the photoclinometric profile. This phenomenon does not occur for the 108-km crater, indicating that at larger sizes the finite element mesh is more sensitive to short-wavelength effects. The disparity between the plateau-like appearance of the resulting profile and the more rounded appearance of the Neith profile could potentially be reconciled by the notion that Neith possesses a wider pit: possibly, the link between thermal anomaly size and pit size breaks for larger pits. A smaller-sized thermal anomaly would uplift the terrain at a very localized region in the center of the pit preferentially, tapering in height with increasing radial distance, and presumably resembling a dome more than a plateau. Future simulations could explore this hypothesis. For high heat fluxes (Figure 10b), all topography is relaxed within 10 Myr, suggesting that very large pit craters with wide central pits that formed under ancient heat fluxes would not form dome craters. However, while the result is physical, it pushes the limits of resolution of our finite element mesh due to the reduced thickness of the lithosphere: the lithosphere has a fixed thickness for a given heat flux, so proportionately, it is a thinner layer relative to the element size for the larger craters. The lithosphere is still resolved for this case, but there is an upper limit to how large the crater can be using this discretization scheme. This case could have implications for penepalimpsest and palimpsest formation, but these crater classes are beyond the scope of this study as their formation may involve penetration into a liquid water layer in the near-surface (Moore et al., 2022).

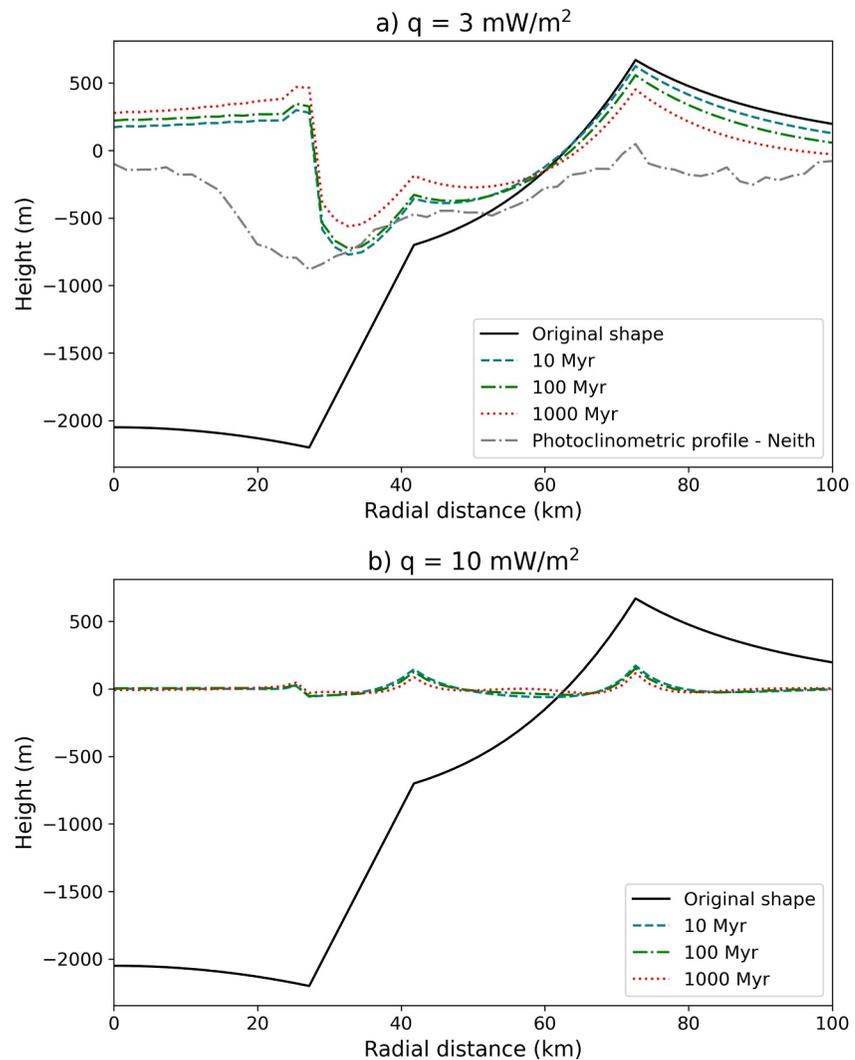

**Figure 10.** Topographic profiles of a relaxing 145-km pit crater under a heat flux of (a) 3 mW m$^{-2}$ and (b) 10 mW m$^{-2}$; a thermal anomaly is included for both cases. Topography is not retained at high heat fluxes for craters of this size.

### 3.2. Simulations Over the Pit-Crater Size Range (D ~35–60 km)

For craters with diameters of 40 km, our simulations show no dome formation regardless of heat flux (Figure 11). Under low heat flux, little relaxation is observed (cf. Bland & Bray, 2024). Under high heat flux, there is overall relaxation of the crater, but no dome is formed. Both cases include the thermal anomaly left over from the impact.

### 3.3. Simulations Over the Overlapping Pit Crater/Dome Crater Size Range

Near the morphological transition diameter between pit craters and dome craters (~60 km), both pit and dome craters are observed on Ganymede and Callisto. We explore the conditions that could lead to one or the other at crater diameters of 58 km. A thermal anomaly left over from the impact below the pit is included in all cases.

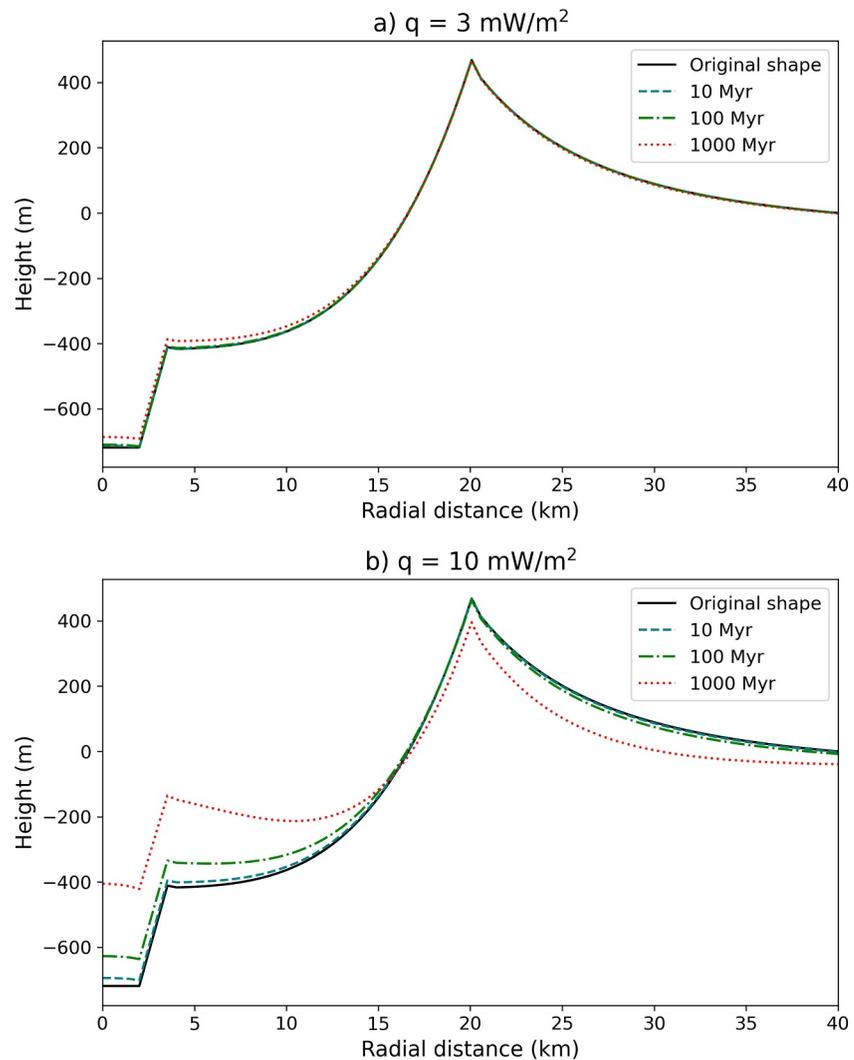

**Figure 11.** Topographic profiles of a relaxing 40-km pit crater under a heat flux of (a) 3 mW m$^{-2}$ and (b) 10 mW m$^{-2}$. Under low heat flux, little relaxation is observed, and no dome is formed. Under high heat flux, there is overall relaxation of the crater, but no dome is formed.

At 58 km in size, Ganymede has a pit crater with a relatively narrow central pit at (20.3°S, 304.3°E), and a dome crater with a wider central pit at (13.5°S, 25.3°E) as shown in Figure 12.

Our simulation results for the 58-km craters are shown in Figure 13. For the narrow-pit case (Figure 13a), our simulations show that the crater stays as a pit crater under both low heat flux (3 mW m$^{-2}$) and high heat flux (10 mW m$^{-2}$) conditions. For the wider-pit case (Figure 13b), the crater stays as a pit under low heat flux but turns into a dome under high heat flux conditions. This suggests that, within the size range where pits and domes coexist, a pit crater is prone to evolve into a dome if its central feature is sufficiently broad and if the crater was subject to high heat flux at the time of formation. Notably, the crater in Figure 12b is inferred to be geologically older than the one in Figure 12a due to the number of superposed craters and its general degradation. This is in line with the premise of higher ancient heat fluxes, lending additional support to our simulation results. The preference for a wider central feature in the evolution of craters into domes is consistent with the dependence of relaxation time on wavelength (e.g., Dombard & McKinnon, 2006a; Scott, 1967): longer wavelengths relax faster due to the larger amounts of displaced mass involved that lead to larger driving forces. In addition, the thermal anomaly below the wider-pit case is spatially larger, contributing to shorter relaxation times.

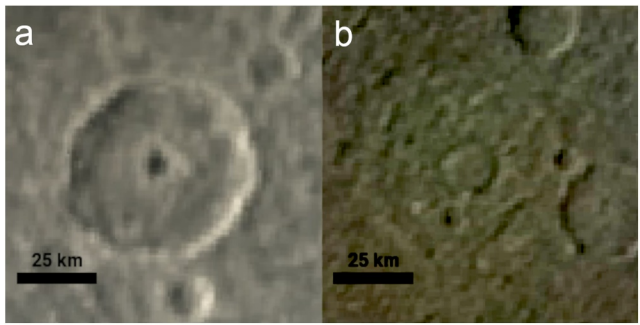

**Figure 12.** Voyager image of pit and dome craters on Ganymede with diameters of 58 km. (a) Unnamed central pit crater with a relatively narrow central pit (20.3°S, 304.3°E). (b) Unnamed central dome crater with a relatively wide central pit (13.5°S, 25.3°E).

### 3.4. Dome as Initial Topography

We altered the starting shape to that of a dome crater instead of a pit crater to test the potential effects if dome craters were formed during an impact, as previous hypotheses for dome formation have suggested. In our simulation, we include the residual heat from the impact and model the evolution of a 108-km dome crater over the span of a billion years under a background heat flux of 3 mW m$^{-2}$. We find that if a dome was formed initially during the impact, it would retain a dome shape. Utilizing a simplified Osiris topographic profile as the starting topography, we find that the evolution up to 10 Myr would largely coincide with the initial profile (see Figure 15). After 100 Myr, the profiles diverge from the photoclinometric data, which may not be relevant for a young crater like Osiris, but could be relevant for older dome craters. This could undermine the hypothesis that dome craters could form on impact and remain this shape after relaxation.

### 3.5. Simulation Result Summary

A summary of all the relevant simulations with the geometries, simulation parameters, and outcomes (whether or not domes formed) is provided in Table 1.

## 4. Discussion

We show that dome craters on Ganymede and Callisto can form via topographic relaxation of pit craters in short timescales of 10 Myr. Topographic relaxation restores a flat surface in response to a perturbation in the deviatoric stress state of the area induced by crater topography. Ice flows downwards from the rim where there is excess mass, and upwards from the crater depression and especially the pit where there is missing mass. High temperatures located right below the central pit, caused by remnant impact heat, soften the ice and allow it to flow more readily than its surroundings. As a result, the upward flow of material is channeled below the pit, and this preferential uplift of material below the pit causes a dome to emerge. Our results can be compared to Bland and Bray (2024), who performed similar simulations without remnant impact heat and obtained some minor upbowing of the pit floor with amplitudes insufficient to explain the observed domes (cf. our simulations without remnant heat [Figure 7]).

Longer wavelength topography relaxes faster than shorter wavelength topography of the same amplitude. Because relaxation is driven by deviatoric stresses that result from the total weight of the topography (the topographic pressure integrated over the horizontal baseline), longer wavelength topography means more driving stresses and therefore faster relaxation rates (Melosh, 1989; Scott, 1967). This phenomenon partly explains the transition we observe from pit craters that do not relax into domes at small crater sizes of 40 km (Figure 11) to pit craters that relax into domes at larger crater sizes of 108 and 145 km in diameter (Figures 9 and 10). We further observe the importance of the topographic wavelength when we explore crater sizes around the morphological transition between the pit crater class and the dome crater class (~60 km). Craters of the same size and depth relax differently for different widths of the central pit floor: for high heat fluxes of 10 mW m$^{-2}$, only wide central pits form domes.

Indeed, another controlling factor in relaxation is the background heat flux (cf. Bland & Bray, 2024; Dombard et al., 2007). Background heat flux creates a geothermal gradient that determines the temperature of the ice below the crater. If the ice surrounding the thermal anomaly is warmer, it can flow more readily due to its lower viscosity, accommodating the preferential uplift of the pocket of high-temperature ice. That is, for the pocket of softer ice to flow, the surrounding material must be able to flow out of the way. The background heat flux is what ultimately determines whether the ice surrounding the thermal anomaly will be able to accommodate the uplift. This mechanism also plays a role in the size transition: craters exhibiting larger pits also possess more extensive thermal anomalies beneath them, which infiltrate into warmer and more mobile subsurface material. Consequently, larger pit craters are more prone to relax into domes. Based on our results for the 58 km crater (Figure 13), background heat flux can determine whether a pit crater stays in that form or whether a dome emerges. A pit crater of that size with a wide central pit relaxes into a dome under high background heat flux conditions (Figure 13d) but remains as a pit crater under low background heat flux (Figure 13c). This could have implications for crater age as older craters are more likely to have been subjected to higher background heat flux.

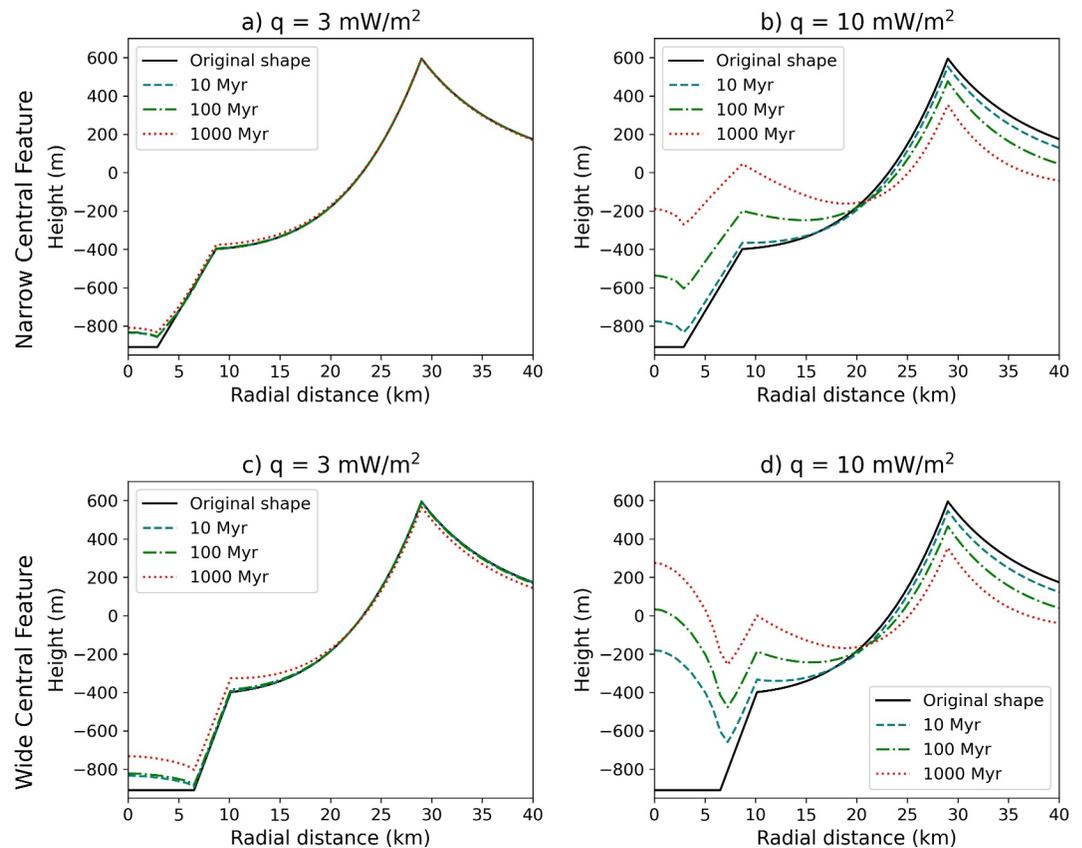

**Figure 13.** (a) Topographic profiles of a relaxing 58-km pit crater with a narrow central pit at low heat flux of 3 mW m$^{-2}$, (b) high heat flux of 10 mW m$^{-2}$, (c) a wider central pit at low heat flux of 3 mW m$^{-2}$ and (d) high heat flux of 10 mW m$^{-2}$. A dome is formed only under high heat flux with a wider central feature.

Background heat flux can also determine the final morphology of the dome crater (cf. Figures 9 and 10). In the case of the 108 km crater, we obtain higher domes for higher heat flux because of the higher relaxation rates (Figure 9b). In the case of the 145 km crater, the compounded effects of the longer-wavelength topography, thermal anomaly spatial extent, and high heat flux lead to even higher rates of relaxation that result in the loss of most topography (Figure 10b), with possible implications for penepalimpsest or palimpsest formation. However, the formation of penepalimpsests and palimpsests may require a preexisting liquid layer in the near-surface based on their distinct morphology (Moore et al., 2022). Because the mechanism for dome formation requires solid-state flow of ice, domes would not form if the near-surface was liquid, as may have occurred under higher ancient heat fluxes, possibly leading instead to the origin of penepalimpsests and palimpsests. Therefore, by analyzing the current morphology of dome craters, we can deduce the background heat flux conditions at the time of their formation, making these structures a record of the planetary thermal history.

As stated above, pit craters with wide central features near the pit-to-dome size transition form domes in our simulations only under high heat flux. This finding suggests that the heat flux when these domes formed was higher. Two observed trends support these simulation results: domes tend to occur in the pit craters with larger central pits (Schenk, 1991) and dome craters with large central features apparently occur preferentially in older, dark terrain (Moore & Malin, 1988). Thus, it appears that pit craters with wider central features both tend to be older and to present domes, which could mean that they were formed under higher ancient heat fluxes. There is still the question of why wide central features would form preferentially under high heat flux. One explanation could be that higher heat fluxes could lead to a warmer central region and larger pits due to slower heat diffusion in the solidifying meltwater pocket (see below for our hypothesis on pit formation).

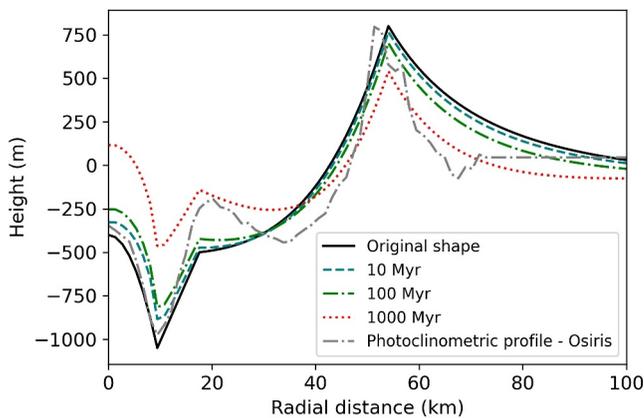

**Figure 14.** Topographic profiles of a relaxing 108-km dome crater under a heat flux of 3 mW m$^{-2}$. The evolution up to 10 Myr closely resembles the topographic profile of Osiris.

Our dome crater profiles largely match the profiles for a range of dome craters from Osiris to Neith measured from the photoclinometric elevation models of White et al. (2022). We obtain a shallow crater with a dome sitting on a rimmed pit, which matches the general morphology of dome craters. The formation of the pit rim occurs as the terrain upbows for all craters that relax into domes in our simulations. However, for the 40-km pit craters that do not relax into domes and stay as pits, pit rims only form for high heat fluxes (Figure 11b), suggesting that pit rims could still need a different formation mechanism apart from relaxation.

Topographic relaxation is the process of restoration of hydrostatic equilibrium which ultimately would result in a flat surface. Achieving a flat surface equilibrium state depends on how much time passes and how easily the material can flow (viscosity). The dome topography reached in these simulations is a quasi-equilibrium state where the driving stresses and the temperatures have decreased enough to diminish the rate of relaxation considerably. The quasi-equilibrium state is retained in time because the thermal anomaly largely dissipates after 10 Myr, and the driving stresses decrease as the topography readjusts. In our simulations, we observe that most deformation happens in the first 10 Myr when the thermal anomaly is still present. Subsequent to this period, the rate of deformation exhibits a notable decrease, where an additional billion years contribute only marginally to further deformation.

### 4.1. Effects of Initial Geometry

Small changes in the geometry while preserving topographic wavelength did not significantly affect the results. Halving the pit depth of the Osiris-sized pit crater rendered similar topographic profiles as the case shown in Figure 7. This finding further suggests that the results are not an artifact of a particular initial geometry but a quasi-equilibrium state reached as the material gets rid of the driving stresses. For shallower pits, the driving stresses are reduced, but the amount of relaxation required to attain a quasi-equilibrium state is correspondingly diminished. Simulations where the pit rim is included in the initial shape and the pit floor is flat were also conducted (Caussi et al., 2020), rendering the same general results.

Using a dome crater profile similar to that of present-day Osiris as the starting point for these simulations also yields a dome crater profile similar to present-day Osiris (Figure 14). This finding is consistent with the quasi-equilibrium idea, but it also means that if dome craters were formed using a different process, their topography would be retained. Thus, topographic relaxation acting alone can explain dome formation from a pit but does not rule out alternative hypotheses. However, the simulations reveal a discrepancy between the topographic profiles obtained after 10 Myr and the initial shape of the dome crater, which is based on present-day Osiris. This difference suggests that if the simulation continued beyond this time frame, the dome would grow taller than what is observed in Osiris. This finding introduces the possibility that for a dome crater older than Osiris, there might be a mismatch between its profile and the results from our simulations, undermining the alternative hypothesis that dome craters could start as and remain dome-shaped upon relaxation.

### 4.2. Mechanism for Pit Formation

We have shown how dome craters can evolve from central pit craters, but the question of where the pits come from in the first place remains. One clue may come from albedo contrasts. Domes have been hypothesized to be composed of cleaner ice relative to their surroundings because of their higher albedos (Schenk, 1993). However, a more recent study (Lucchetti et al., 2023) did not find those compositional differences upon analysis of Melkart crater. Our hypothesis does not conflict with the idea of a different composition in the dome material because any initial process such as extrusion can create a small "seed dome" made of cleaner ice, and our model would then explain the final dome heights achieved. A small seed dome is included on the pit floor of the initial topography in our simulations to account for these possible differences in texture or color in the dome caused by any initial process.

A possible initial process causing the higher albedos (and the fractures on the domes) could occur during pit formation via melt drainage. Pit formation via melt drainage has been proposed previously and consists of impact melt draining into the subsurface while leaving a void that originates the pit (e.g., Elder et al., 2012). Here, we build upon that hypothesis to propose a possible explanation for the origin of pits and for the differences in color and texture of the domes (Figure 15). Hydrocode simulations (Korycansky, Caussi, et al., 2022) showed that impacts into relatively warm surfaces of large icy moons such as Ganymede and Callisto produce a damaged region, and a large pool of meltwater below the crater center, which precludes the formation of a central peak (Figure 15a, cf. Figure 6). Other icy moons, such as the Saturnian satellites, do not exhibit central pits but present central peaks instead. Ganymede and Callisto are two of the largest icy worlds, and are also relatively close to the Sun; we speculate that these conditions might explain the differences. On a larger-sized moon like Titan, conditions may allow for the formation of a larger pocket of melt, but no central pits are observed either. Future hydrocode simulations could explore this specific question.

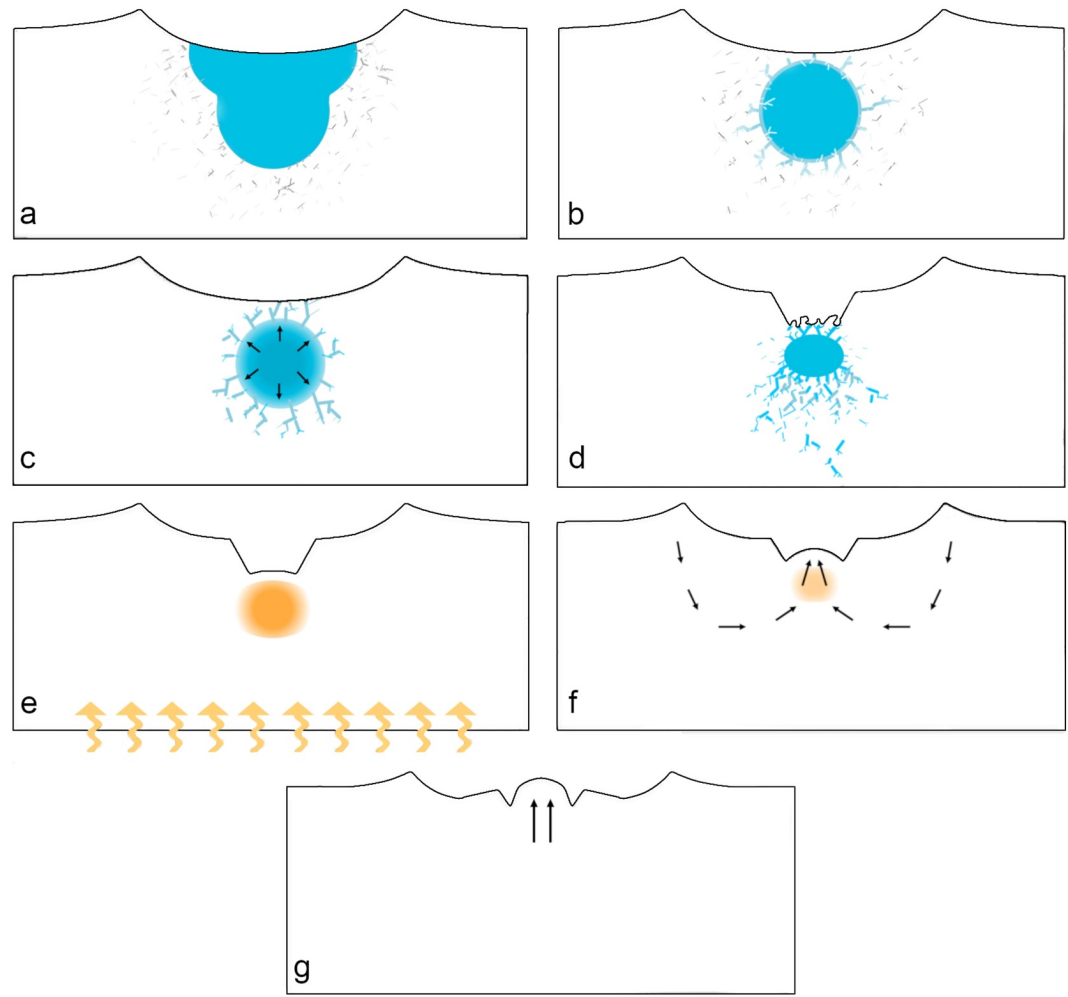

**Figure 15.** Scenario for pit and dome crater formation. (a) Impact into icy target damages region and creates a large pool of meltwater (liquid water shown in blue). (b) Melt pool cooling and freezing seals cracks in the damaged region, and surface freezing provides a roof. (c) Volume expansion during solidification stresses and cracks the surrounding ice. (d) Seal on melt pocket breaks and water drains into the damaged region. Roof collapses and pit forms. (e) Melt solidifies, leaving a warm pocket of ice below the pit, indicated by an orange oval. (f) Ice creeps, relaxing stresses induced by topography. Flow enhanced in the central region due to low viscosity. (g) Preferential uplift in the central region causes the dome to emerge from the pit.

Some of the water could drain into cracks in the damaged region and seal them upon refreezing (Figure 15b). This process, along with surface freezing of the melt pool, would create a subsurface pocket of meltwater trapped in a confined space that is cooling down. As the confined melt cools, it would solidify and expand, stressing the surrounding ice (Figure 15c). With sufficient stress, the seal on the melt pocket could break, leading to liquid drainage and subsequent roof collapse, leaving a pit (Figure 15d). More details on the hydrocode simulations and the evolution of the melt pool, including time scales, will be presented in future work. For now, the work of Korycansky, Caussi, et al. (2022) establishes the basic plausibility.

Some remaining melt could extrude through the collapsed roof, generating the albedo differences upon solidification. The small seed dome included in our simulations could represent this extrusion of trapped melt. A similar idea was explored in Quick et al. (2019), where a cryomagma chamber below the Occator dome crater in Ceres solidified while stressing the region and extruded liquid to the surface to form the dome.

Subsequently, all the remaining water below the pit would freeze, and that would be the starting point of our simulations. The warm, refrozen ice below the pit (Figure 15e) with lower material viscosities has been shown here to cause a dome to emerge within the pit as the crater relaxes over time scales of ∼10 Myr (Figures 15f and 15g).

Supporting this hypothesis, we find that the most accurate match to the Osiris photoclinometric profile is achieved when the thermal anomaly is strictly confined beneath the pit floor. This outcome aligns with the proposed pit formation process discussed earlier. If the pit is the result of partial drainage of the melt pocket, then logically, the melt pocket's spatial extent must mirror that of the pit floor to some degree.

Alternatively, topographic relaxation alone could explain the albedo and texture differences. Fractures observed in the domes can be caused by dome uplift and extension during relaxation, with the cold layer of ice in the surface behaving brittly. In our simulations, the differential stresses can reach maximum values at the surface of the dome of up to ∼5 MPa. The tensile strength of ice is of comparable magnitude (Petrovic, 2003), and could be exceeded causing tensile fractures in the dome. Similarly, it is also possible that topographic relaxation could by itself cause an albedo difference. Topographic relaxation uplifts ice at depth into the center of the pit; if the dark component in the ice is a thin deposit at the surface, any movement upwards could disturb the deposit and brighten the area by exposing cleaner ice. Therefore, a small seed dome generated using a different process is not strictly necessary to account for the brightness. Future high-resolution observations will help resolve the origin of these possible albedo differences.

## 5. Conclusion

Our simulations show that dome craters on Ganymede and Callisto may form by topographic relaxation of pit craters aided by remnant heat from the impact within a timescale of ∼10 Myr. The rapidity of dome formation suggests that even young dome craters such as Osiris on Ganymede may be the consequence of relaxation from an initial central pit crater morphology.

Topographic relaxation occurs in response to a perturbation of the deviatoric stress state of the area induced by crater topography. Gravity acts as a restorative force, causing ice to flow downward from the rim, where there is excess mass, and upward from the crater depression where there is a deficit of mass. High temperatures located immediately below the central pit, caused by remnant impact heat, soften the ice and allow it to flow more readily. Consequently, the upward flow of material is channeled below the pit, causing the material below the pit to uplift preferentially and form a dome.

Our model for dome formation reproduces the size transition between the pit crater class and the dome crater class observed on Ganymede and Callisto. Pit craters are the predominant class for sizes up to ∼60 km, above which domes become predominant. Our simulations show that pit craters undergo relaxation but do not turn into dome craters at sizes below ∼60 km. For sizes above ∼60 km, a dome emerges from the pits as the craters relax. At the transitional size range, the specific conditions and geometries determine whether the crater stays as a pit crater or emerges as a dome. The size transition between crater classes is consistent with the wavelength dependency of topographic relaxation, where wider features relax more rapidly than narrower features. Larger pit craters above the size threshold undergo faster relaxation, resulting in the final dome morphology, as opposed to smaller pit craters where the pace of relaxation is sluggish, and no dome emerges.

Background heat flux also plays a role in relaxation by softening the ice surrounding the plug of high temperatures left over from the impact. The preferential uplift of the region below the pit must be accommodated by the movement of the surrounding ice; higher background heat fluxes facilitate this movement. Because the mechanism for dome formation requires solid-state flow of ice, domes would not form if the near-surface was liquid, as may have occurred under higher ancient heat fluxes, possibly leading instead to the origin of penepalimpsests and palimpsests. Given the timescales involved and the dependence on heat flux, our dome formation model could be utilized to constrain the thermal histories and crater ages of these moons.

The results of our simulations align well with the digital elevation models of dome craters obtained through photoclinometry. The dome heights predicted by our model for Osiris and Neith craters on Ganymede correspond reasonably to the digital elevation models. This result indicates that topographic relaxation alone can account for the final dome heights, provided that the thermal anomaly from the impact is included in the model.

While our topographic relaxation model provides an explanation for the observed dome morphologies, it may not fully account for the textures observed on domes, such as fracture networks and higher albedos. The uplift process during relaxation can explain the fractures as the floor of the pit experiences tension, reaching stresses comparable to the tensile strength of ice. However, an additional initial process might be required to explain the albedo difference. Central pit craters have been hypothesized to be formed by impact melt drainage that leaves a cavity. We build on that hypothesis to suggest that some water might be trapped during that process and extruded upwards to cause the higher albedo. The pressure buildup that leads to the extrusion might also offer an explanation for the presence of fractures.

## Data Availability Statement

The commercial finite element package Marc is available for use with academic pricing (Hexagon, 2023). The output text files for each simulation described in Table 1 are available on Zenodo (Caussi et al., 2024).

## Acknowledgments

Funding was provided by the NASA Solar System Workings Program award 80NSSC19K0551. We thank our anonymous reviewers for their insightful contributions to our manuscript.